\documentclass[epj]{webofc}
\usepackage[varg]{txfonts}   
\begin{document}
\title{Search for $Z'$, vacuum (in)stability and hints of high-energy structures\footnote{Presented at $QCD@Work2016$, 27-30 June 2016, Martina Franca, Italy}}

\author{\firstname{Elena} \lastname{Accomando}\inst{1}\fnsep\thanks{\email{e.accomando@soton.ac.uk}} \and
        \firstname{Claudio} \lastname{Corian\`o}\inst{2}\fnsep\thanks{\email{claudio.coriano@le.infn.it}} \and
        \firstname{Luigi} \lastname{Delle Rose}\inst{1,3}\fnsep\thanks{\email{l.delle-rose@soton.ac.uk}} \and
        \firstname{Juri} \lastname{Fiaschi}\inst{1}\fnsep\thanks{\email{juri.fiaschi@soton.ac.uk}} \and
        \firstname{Carlo} \lastname{Marzo}\inst{2}\fnsep\thanks{Speaker. \email{carlo.marzo@le.infn.it}} \and
        \firstname{Stefano} \lastname{Moretti}\inst{1}\fnsep\thanks{\email{s.moretti@soton.ac.uk}}
}

\institute{School of Physics and Astronomy, University of Southampton, Highfield, Southampton SO17 1BJ, UK
\and
	           Dipartimento di Matematica e Fisica Ennio De Giorgi, Universit\`a del Salento and INFN-Lecce, \\ Via Arnesano, 73100 Lecce, IT 
\and
           Dept. of Particle Physics, Rutherford Appleton Laboratory, Chilton, Didcot, OX11 0QX, UK
          }

\abstract{%
We study the high-energy behaviour of a class of anomaly-free abelian extensions of the Standard Model. We focus on the interplay among the 
phenomenological characterisation of the model and the use of precise renormalisation group methods.  
Using as boundary conditions regions of the parameter space at the verge of current LHC probe,
interesting unification patterns emerge linked to thresholds belonging to a SO(10) grand unification theory (GUT).
We stress how the evolution of the mixing between the two abelian factors may provide a valuable tool to address the candidate high-energy embedding. 
The emerging unification scenarios are then challenged to be perturbative and to allow for a stable vacuum. 
}
\maketitle
\section{Introduction}
\label{intro}
An extra U(1)' gauge symmetry is a common presence in many attempts to go beyond the Standard Model (SM).
It represents, from a low-energy perspective, the simplest extension that can be attached to the SM gauge group,
at the same time, from the opposite high-energy point of view, an extra abelian factor is almost an unavoidable leftover from the breaking of many GUT scenarios \cite{Langacker:2008yv}.\\
If we adopt a (grand) unification paradigm, it is therefore feasible that a regime ruled by the gauge structure SU(3) $\otimes$ SU(2) $\otimes$ U(1$)_Y$ $\otimes$ U(1)'
could populate the sequence of effective descriptions scaling from the GUT energy, before breaking into the SM one.
The last step may be triggered by the non-trivial vacuum expectation of a scalar field $\chi$, that is, consequently, required to be SM-singlet.
If such U(1)' breaking is realised at the TeV scale, then there are realistic prospects of an interesting interplay with the current LHC probe, 
the precise traits of such phenomenological characterisation being dictated by the extended matter content (\cite{Emam:2007dy,Basso:2008iv,Basso:2010yz,Accomando:2013sfa,Accomando:2015cfa}). Beyond the scalar sector, where a SM-singlet accounts for the extra U(1)' breaking,
and the neutral vector $Z'$, to accomplish gauge invariance, one extra fermion per generation is needed to cancel gauge and gravitational anomalies in a minimal way.
This scenario has been the subject of a recent up-to-date investigation \cite{Accomando:2016sge} where we exploited the bounds and the discovery potential of current and forthcoming collider searches. 
The more promising regions of the allowed parameter space have supplied the boundary conditions for a Next-to-Leading-Order (NLO) vacuum stability analysis, that we performed extrapolating 
the model to higher energies with two-loop $\beta$ functions. As a result, the explored regions have been labeled with the maximal energy scale up to which they 
would provide a coherent (stable and perturbative) extrapolation of the model. \\ 
The role of the Renormalisation Group (RG) extrapolation does not exhaust its insight power with the stability analysis. As we will illustrate, for the particular case of our minimal SM $\otimes$ U(1)'
regime, the RG may draw clear indications also about the high-energy regime that is expected to take place. 
In a combined effort, all the phenomenological and formal aspects of this analysis will contribute to unveil a consistent link between the low-energy model characterisation,
and a stable, perturbative, ultraviolet (UV) completion. 
\section{Structure of the model and constraints from current collider probe}
\label{sec-1}
\begin{figure}[h]
\centering
\includegraphics[width=4cm,clip]{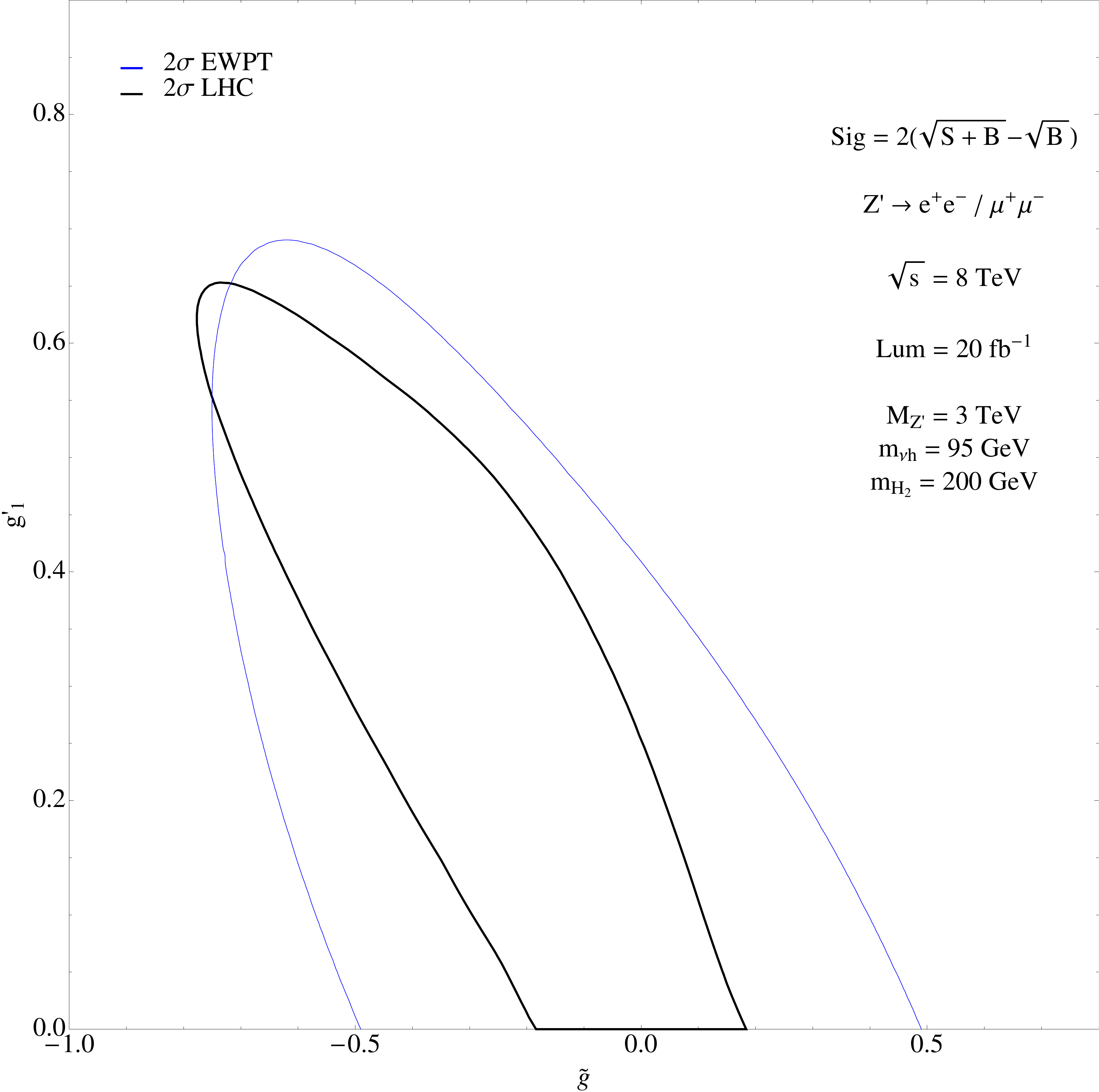} \qquad \qquad
\includegraphics[width=4cm,clip]{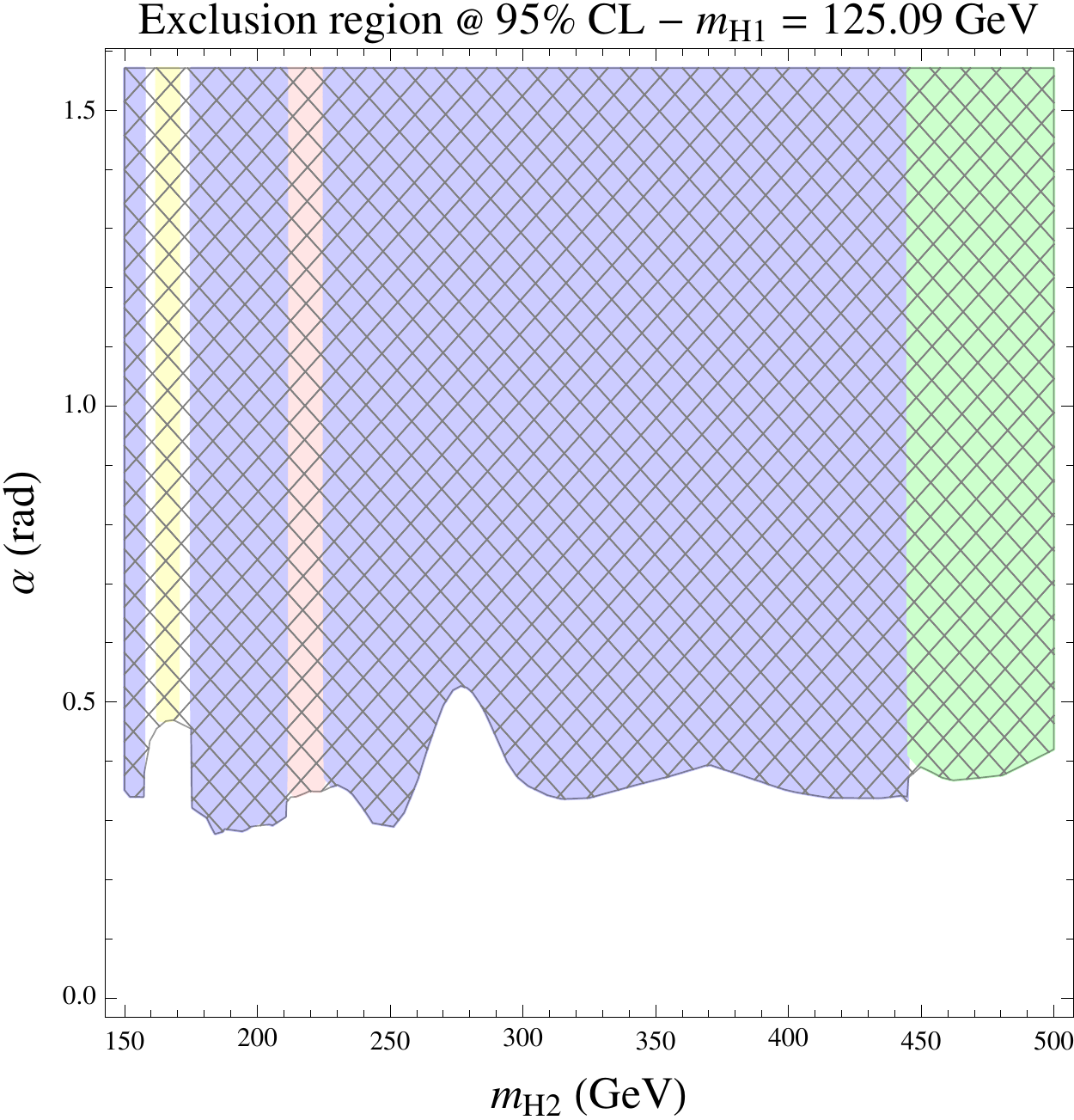}
\caption{(a) EWPTs and LHC bounds for the case $M_{Z'} = 3$ TeV (b) Excluded region in the $(m_{H2}, \alpha)$ parameter space for $m_{\nu_h} = 95$ GeV from \texttt{HiggsBounds}.}
\label{fig-1}     
\end{figure}

The class of models encompassed in \cite{Accomando:2016sge} adds, to the SM field content, a massive neutral gauge boson
plus a scalar and three extra fermions, all transforming trivially under the SM gauge group. 
This extended spectrum is naturally introduced to minimally account for gauge invariance, anomaly freedom and the requirement of a massive $Z'$. 
We notice, as a valuable consequence of the previous setting, that the presence of a new abelian factor has forced the introduction of states that
complete the 16-dimensional representation of SO(10) which is a further motivation to explore the possible UV fate of such model. 
Anomaly cancellation also rules the possible U(1)' charges, leaving to the ratio of just two parameters the definition of the allowed charge assignments.
In a low-energy investigation the common choice is to highlight the Hypercharge operator $Y$, so that the generator of the extra U(1)' is constrained to the form 
$Y' = \left(B - L\right) + \left(\tilde{g}/g'_1\right)Y$,
where B and L are, respectively, the Baryon and the Lepton number of the fields. The overall gauge strength $g'_1$ and the 
\emph{mixing} $\tilde{g}$ rule, therefore, the content of the $B-L$ and the Hypercharge operators in $Y'$. \\
The renormalisable interactions that arise from the extended field content can promptly realise a Type-I seesaw mechanism to account for neutrino masses. 
The SM Yukawa sector is in fact supplemented by the new contribution
\begin{equation}
\mathcal{L}' =  - Y_\nu \, \overline{L} \, \tilde H \, {\nu_R}   - Y_N \, \overline{(\nu_R)^c} \, {\nu_R} \, \chi
\end{equation}
 that, after the sequential breaking 
of the U(1)' and of the Electro-Weak (EW) symmetry itself, provides the wanted massive state with the typical seesaw hierarchy. The scalar sector, instead, is extended to the form 
\begin{equation}
V(H,\chi) = m_1^2 H^\dag H + m_2^2 \, \chi^\dag \chi + \lambda_1 (H^\dag H)^2 + \lambda_2 (\chi^\dag \chi)^2 + \lambda_3 (H^\dag H)(\chi^\dag \chi) 
\end{equation}
 where  $H$ is the Higgs doublet state. \\
It comes therefore with no surprise that, when the vacuum expectation value of $\chi$ is at the TeV scale, the model has to deal with the exclusion limits from the more recent collider 
quest. In particular, the past investigations of the $Z'$ have been strongly affected by bounds coming from EW Precision Tests (EWPTs) from LEP2. The first Run
of LHC at 8 TeV and $\mathcal{L} = 20$ fb$^{-1}$ has generated even more stringent bounds at the TeV scale. These can be extracted using a signal-to-background analysis for the Drell-Yan channel.
When translated in constraints in the extended $(g'_1 , \tilde{g})$ plane, the range of the allowed values is visible, for the exemplar case of $M_{Z'} = 3$ TeV,
from fig.~\ref{fig-1}(a). 
Also the extended scalar sector creates numerous chances to reveal and characterise the class of models under study. We have limited the related new parameter space, 
that we parameterised with the new scalar mass $m_{H_2}$ and the mixing angle $\alpha$, considering the bounds from the direct detection probes, and comparing the 
signal produced with the one measured of the discovered Higgs at 125.09 GeV. This analysis has been performed using, respectively, \texttt{HiggsBounds} \cite{arXiv:0811.4169} and \texttt{HiggsSignals} \cite{Bechtle:2013xfa}
tools, and has constrained the $(m_{H_2},\alpha)$ space to the regions of fig.~\ref{fig-1}(b).
\section{The kinetic mixing and the breaking of SO(10) thresholds}
\begin{figure}[h]
\centering
\includegraphics[width=9cm,clip]{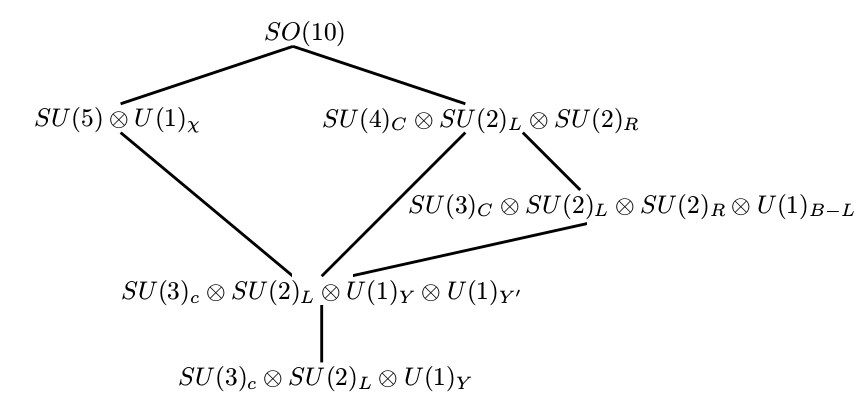} 
\caption{The SO(10) breaking chart illustrating the chains investigated in this work. The cases for the Flipped SU(5) and Pati-Salam required also 
additional unification conditions involving the non-abelian gauge sector.}
\label{fig-3}     
\end{figure}

\begin{figure}[h]
\centering
\includegraphics[width=2.1cm,clip]{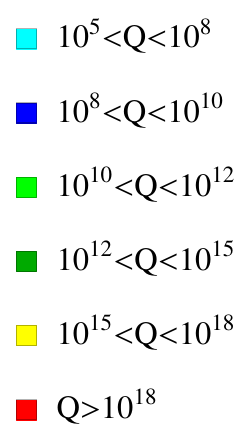} \quad \quad \quad \quad
\includegraphics[width=4cm,clip]{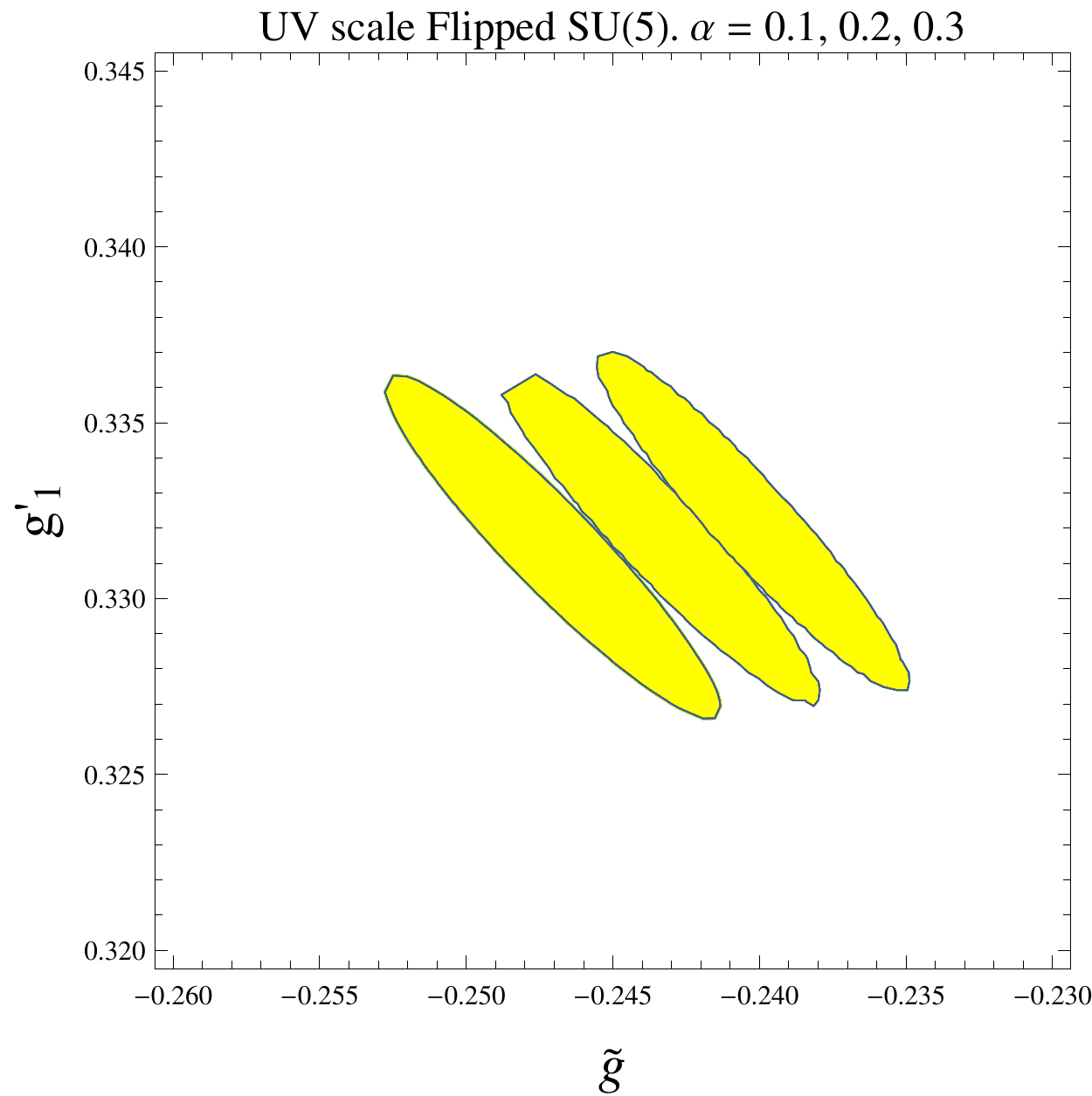}
\includegraphics[width=4cm,clip]{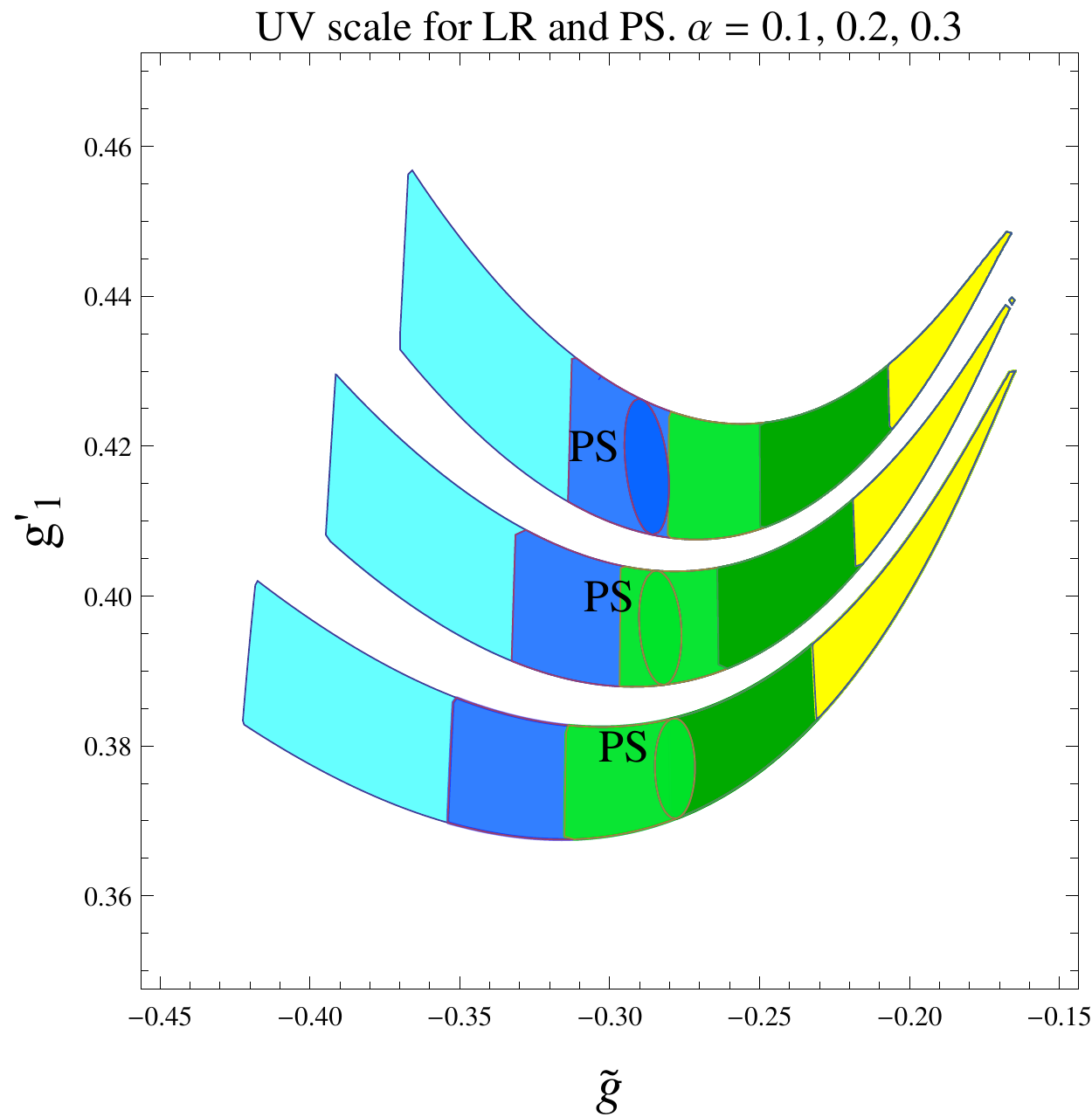}
\caption{(a) When we consider the stability and perturbativity analysis, the colours refer to the different regions defined by the maximum energy up to which
the model is stable and perturbative. The same energy/colour relation is also used for the unification study, referring to regions that fulfil the given unification
requirement. (b) Regions with Flipped SU(5) restoration. (c) Regions with LR and PS restoration. }
\label{fig-4}     
\end{figure}

The presence of multiple abelian factors is a peculiar trait of this class of models. The induced kinetic mixing, absorbed in the contribution of the  coupling $\tilde{g}$
to the covariant derivative
\begin{equation}
\mathcal D_\mu = \partial_\mu + i g_1 \, Y \, B_\mu + i ( \tilde g \, Y + g'_1 \, Y_{B-L} ) B'_\mu + \ldots,
\end{equation}
may shed light, supported by a precise RG inspection, on the UV embedding that precedes the U(1)' regime \cite{delAguila:1995rb,Salvioni:2009mt,Pacholek:2013kca}.\\
The key for this analysis is in the matching of the low and the high-energy generators basis, used to describe the abelian sector. 
In our phenomenological survey, adherence with the SM regime suggested the use of the Hypercharge for one of the two U(1). In turn, within the constraints of
anomaly cancellation, we chose B-L for the other. The mixing would then provide the component of the $Y'$ generator in the Hypercharge direction.  
From a high-energy perspective is more appropriate to work with the basis that is naturally provided by the embedding of the abelian factors in the unifying group. 
For example, a Left-Right (LR) symmetric regime SU(2)$_R \otimes$ U(1)$_{B-L}$, which includes U(1)$_R \otimes$ U(1)$_{B-L}$, would select the corresponding $Y_R$ and $Y_{B-L}$
set of generators. Close to the energy scale of the Left-Right symmetry breaking, the mixing between the $Y_R$ and $Y_{B-L}$ is zero, being protected by the overall 
\emph{non-abelian} gauge symmetry of SU(2)$_R$. It is possible, with the appropriate normalisation, to match our SM-oriented parameters $(g_1,g'_1,\tilde{g})$ with the 
ones corresponding to the (candidate) high-energy basis $(g_R,g_{B-L},\tilde{g}_{R/B-L})$. Therefore, following the RG evolution of $\tilde{g}_{R/B-L}$ in terms 
of $g_1,g'_1$ and $\tilde{g}$, we can recognise, by its zeroing at a given energy, the restoration of a Left-Right symmetry.\\
This analysis can be promptly extended to include thresholds of SO(10) that represent a realistic UV embedding.
These involve, in addition to the Left-Right case discussed, a direct breaking of a Pati-Salam (PS) group into our model, and the flipped SU(5) case (fig.~\ref{fig-3}).  
Choosing as boundary conditions, for the RG extrapolation, benchmarks points inspired by our phenomenological analysis, the parameter space offers regions that, in case 
of discovery, would clearly reveal the presence of one of the previous embeddings (fig.~\ref{fig-4}).
\section{Vacuum stability and Unification patterns}
\begin{figure}[h]
\centering
\includegraphics[width=4cm,clip]{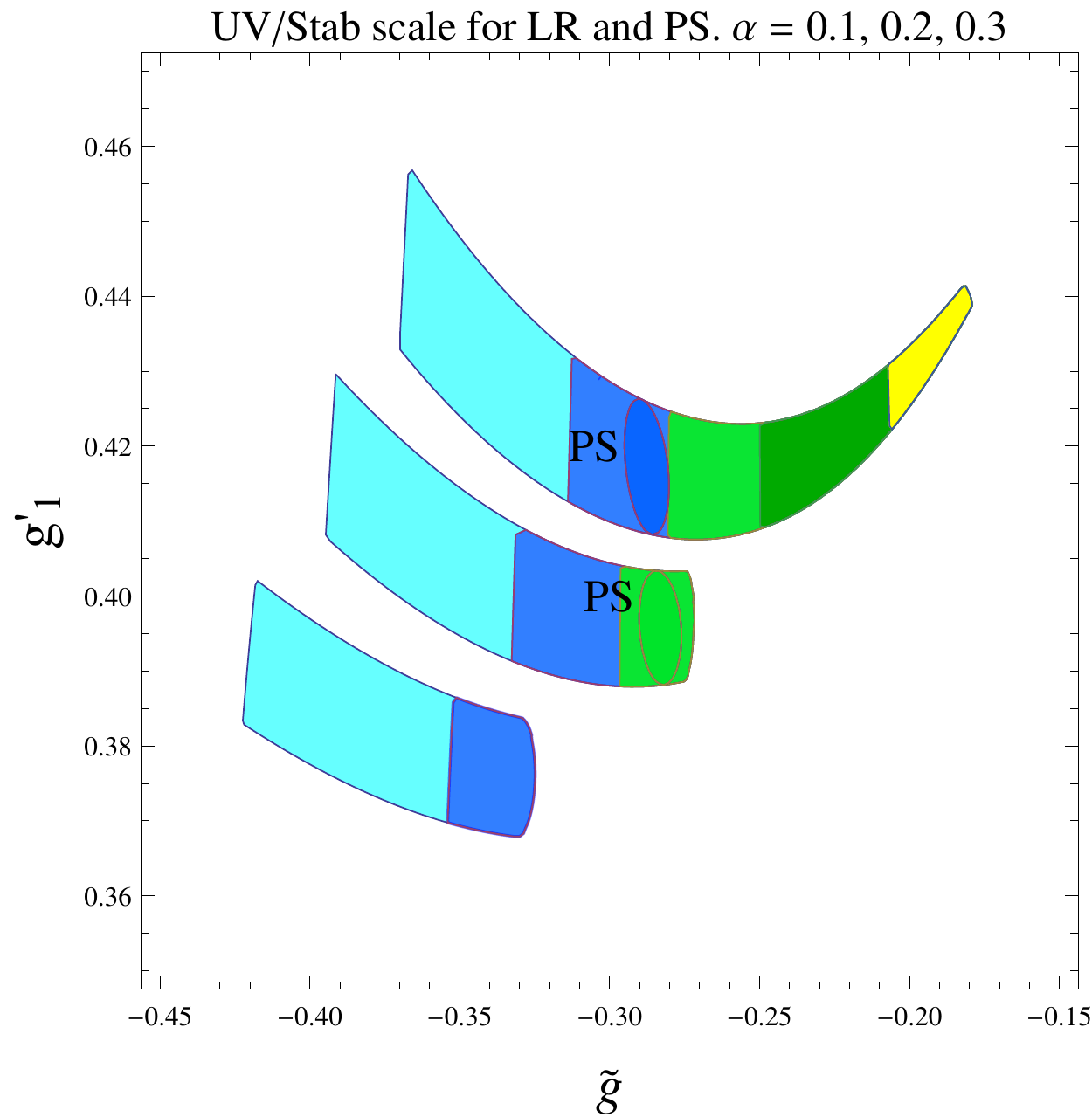} 
\includegraphics[width=4cm,clip]{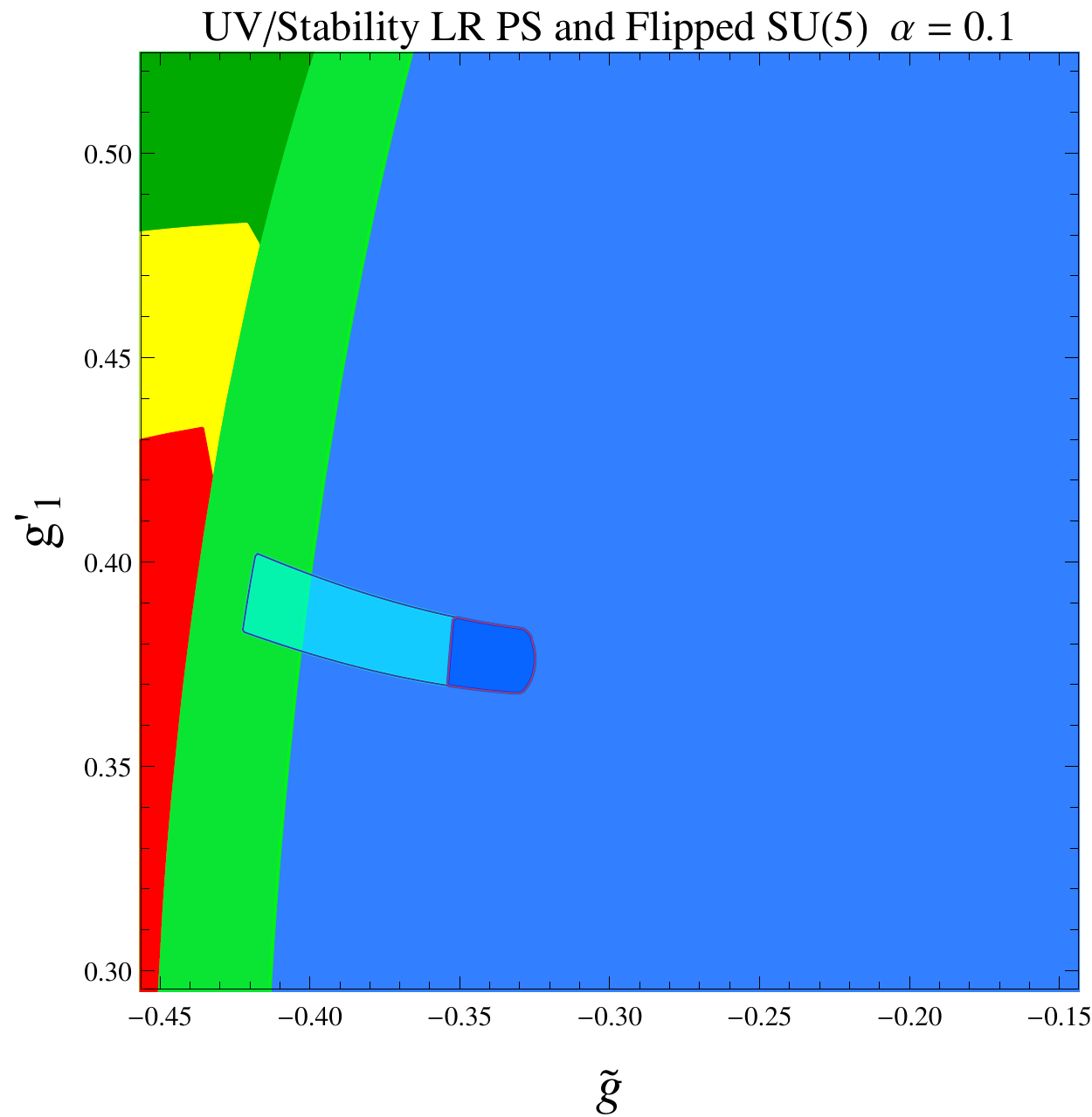}
\includegraphics[width=4cm,clip]{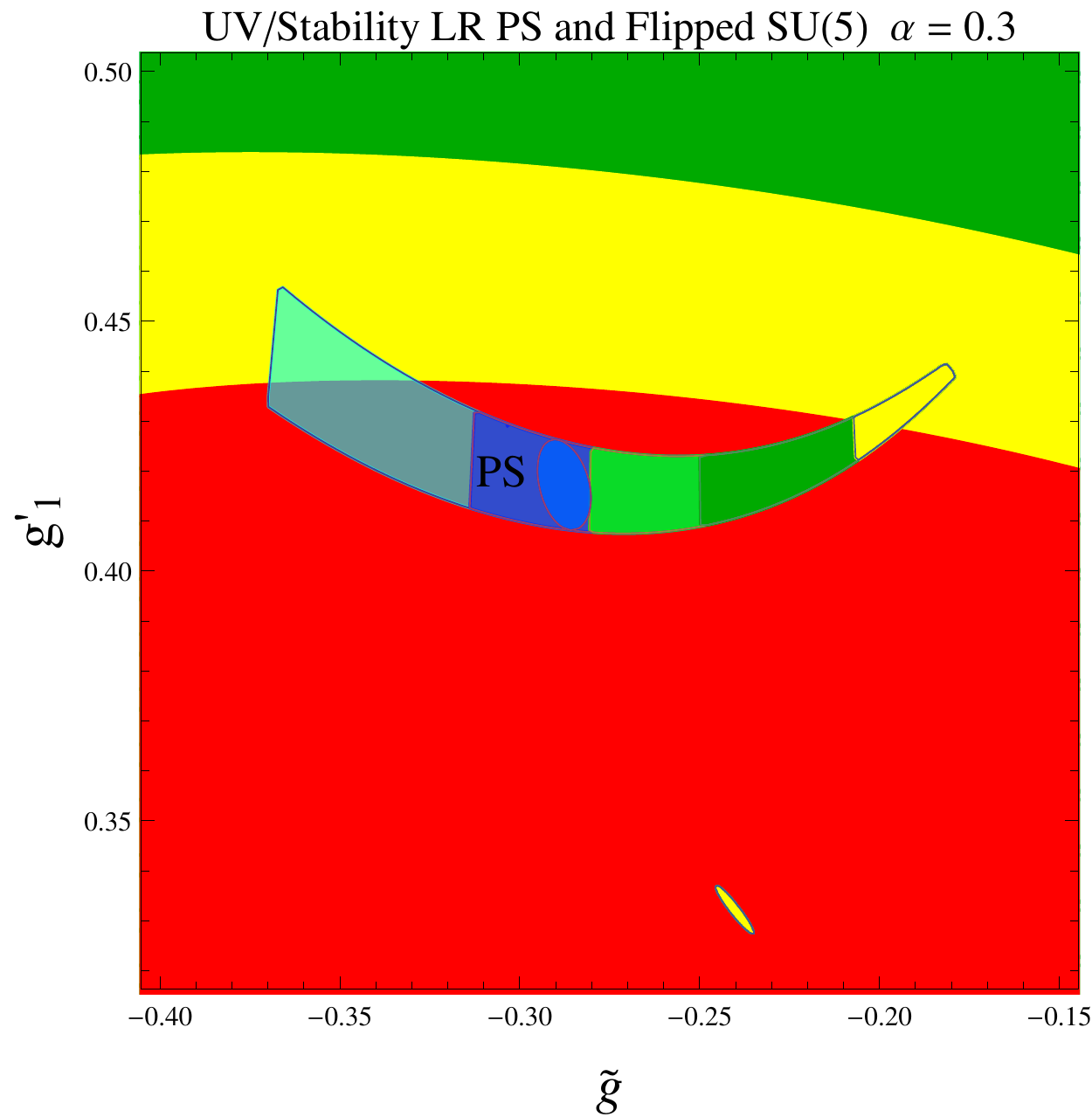}
\caption{(a) The effect of the stability requirement to the LR and PS restoration. The similar analysis for the Flipped SU(5) case would result trivially in the surviving
only of the case with $\alpha = 0.3$. Explicit matching of the stability and perturbativity map with the unification regions. Case  $\alpha = 0.1$ (b) 
and  $\alpha = 0.3$ (c). }
\label{fig-5}     
\end{figure}

The minimal character of this U(1)' extension of the SM makes particularly efficient the use of vacuum stability and perturbativity as constraining requirements
to shape the viable parameter space \cite{Basso:2010jm,Coriano:2014mpa,Coriano:2015sea}.
The vacuum stability is addressed asking for the extended scalar potential to be bounded from below, 
\begin{equation}
\lambda_1 > 0\,, \quad \lambda_2 >0 \,, \quad 4 \lambda_1 \lambda_2 - \lambda_3^2 > 0 \,.
\end{equation}
Together with the perturbativity requirement it is also challenges the viability of a given unification scenario. If we accept the minimal content of the model, then 
a coherent extrapolation asks for the maximum scale of stability and perturbativity to be greater than the one realising a succesful embedding. 
By relying on the analysis presented in \cite{Accomando:2016sge}, we may exploit this further constrains. Our final results (fig.~\ref{fig-5}) give an illustration of 
how the interplay of the tools presented may enrich the forthcoming collider profiling of specific regions of the parameter space. Moreover, in a possible
post-discovery phase, frictions with the measured scenarios would help in outlining the degrees of freedom necessary to recover stability, when
a promising unification of the gauge sector is at hand.  

\section*{Acknowledgments}
E.A., L.D.R., J.F. and S.M. are supported in part through the NExT Institute. C.C. thanks The
Leverhulme Trust for a Visiting Professorship to the University of Southampton, where
part of his work was carried out. The work of L.D.R. has also been supported by the "Angelo
Della Riccia" foundation and a COFUND/STFC Rutherford International Fellowship.

\end{document}